\documentclass[preprint,proceedings]{rmaa}

\usepackage{paralist}

\usepackage{psfrag,color}

\SetYear{2004}
\SetConfTitle{3ra Reunion ADELA}

\title{RR Lyrae Search and Stellar Populations Study in Canis Major: Preliminary Results} 

\author{
  C. E. Mateu,\altaffilmark{1,2} 
  A. K. Vivas,\altaffilmark{1}
  R. Zinn,\altaffilmark{3}
  and L. Miller \altaffilmark{3}}

\altaffiltext{1}{Centro de Investigaciones de Astronom\'\i{}a, M\'erida, Venezuela.}
\altaffiltext{2}{Departamento de F\'\i{}sica, Universidad Sim\'on Bol\'\i{}var, Caracas, Venezuela.}
\altaffiltext{3}{Astronomy Department, Yale University, New Haven, USA.}

\shortauthor{Mateu, et al.}
\shorttitle{RevMexAA(SC)}

\fulladdresses{
\item Cecilia E. Mateu: Centro de Investigaciones de Astronom\'\i{}a (CIDA),
  Apartado Postal 264, M\'erida 5101--A, Venezuela (\email{cmateu@cida.ve}).
\item Lissa Miller: Astronomy Department, Yale University,
  P.O. Box 208101, New Haven, CT 06520--8101, USA (\email{miller@astro.yale.edu}).
\item A. Katherina Vivas: Centro de Investigaciones de Astronom\'\i{}a (CIDA), 
  Apartado Postal 264, M\'erida 5101--A, Venezuela (\email{akvivas@cida.ve}).
\item Robert Zinn: Astronomy Department, Yale University, 
  P.O. Box 208101, New Haven, CT 06520--8101, USA (\email{zinn@astro.yale.edu}).
}

\listofauthors{C. E. Mateu, A. K. Vivas, R. Zinn \& L. Miller}
\indexauthor{Mateu, C. E.}
\indexauthor{Vivas, A. K.}
\indexauthor{Zinn, R.}
\indexauthor{Miller, L.}

\abstract{We present preliminary results of a RR Lyrae star search and stellar populations study performed in the Canis Major overdensity, spanning an area of $8.35$sq deg. The observations were made in R and V bands, with the QUEST camera installed in the 1m Jurgen Stock Telescope, at the Venezuela National Observatory. The resulting Hess diagram shows a possible, but weak, red giant branch and no obvious horizontal branch, red clump or main sequence turnoff. After a multi--epoch photometric search, 6 RR Lyrae stars were confirmed with further observations obtained at the 1.0 and 1.3m telescopes of the SMARTS consortium at CTIO. Of these confirmed RR Lyrae stars, 5 have heliocentric distances between 5 and 7 kpc. Confirmation of their physsical association with the Canis Major system awaits for a study of their radial velocities.}

\resumen{Se presentan resultados preliminares de un estudio de poblaciones estelares y b\'usqueda de estrellas RR Lyrae, realizados en la sobredensidad de Can Mayor, abarcando un \'area de $8.35$gd$^2$ en el cielo. Las observaciones fueron hechas en los filtros R y V, con la c\'amara QUEST instalada en el telescopio Jurgen Stock de 1m ubicado en el Observatorio Astron\'omico Nacional de Venezuela. El diagrama de Hess resultante muestra una posible, pero tenue, rama de las gigantes rojas y no se observan una rama horizontal, red clump o turn-off de la secuencia principal que sean evidentes. Luego de una b\'usqueda fotom\'etrica multi-\'epoca, 6 RR Lyrae fueron confirmadas con observaciones adicionales obtenidas en los telescopios de 1.0 y 1.3m del consorcio SMARTS en CTIO. De dichas RR Lyrae confirmadas, 5 tienen distancias helioc\'entricas entre 5 y 7 kpc. La confirmaci\'on de su asociaci\'on con el sistema de Can Mayor espera por un estudio de sus velocidades radiales.} 

\addkeyword{Galaxies: interaction}
\addkeyword{Galaxy: formation}
\addkeyword{Galaxy: structure}
\addkeyword{Stars: Variable: Other}

\begin{document}
\maketitle

\section{Introduction}

In a recent study of the spatial distribution of 2MASS M giants at low galactic latitudes, Martin et al (2004) found a large stellar overdensity, covering up to 100 sq deg in the sky. The overdensity was shown as a north-south asymmetry in the density of such stars, with respect to the galactic plane. Martin et al interpreted the Canis Major overdensity, as it is now called, to be a dwarf spheroidal galaxy in an advanced stage of disruption due to the Milky Way's tidal forces. In a later study, Momany et al (2004) argued that the Canis Major overdensity could be explained by taking in to account the stellar warp of the galactic disk, which happens to be maximum at a galactic longitude ($l\sim270^o$) lying near the estimated center of Canis Major $(240^o,-8^o)$.  Although further observational evidence has been provided by Bellazzini et al (2004), Mart\'{\i}nez--Delgado et al (2004) and Martin et al (2004b), among others, supporting the dwarf galaxy hypothesis; the nature of the Canis Major overdensity is still controversial.

In order to provide further arguments that would contribute to clarify the nature of Canis Major, we started a large--scale multi--epoch survey in a region near the center of the overdensity, in order to search for RR Lyrae variable stars. An excess of such stars in the region would favor the hypothesis that Canis Major could be a dwarf galaxy with stellar populations similar to other known Milky Way satellites.

\section{Observational Data and Photometry}

The observations were made using the QUEST camera (Baltay et al. 2002) installed on the (1m) Jurgen Stock telescope at the Venezuela National Observatory. Due to the low declinations surveyed, the camera was off the limits for the use of drift-scanning mode, which is the normal mode of operation at this telescope. Hence, it was used in point--and--stare mode, with 2 minute exposures in R and V. The total survey area covers $\sim36$ sq deg. The preliminary results presented here correspond to a smaller area of $8.35$ sq deg. with observations taken in 5--10 epochs between January and March 2004.

\begin{figure}[!th]\centering
\includegraphics[bb=40 40 600 600,clip,width=\columnwidth]{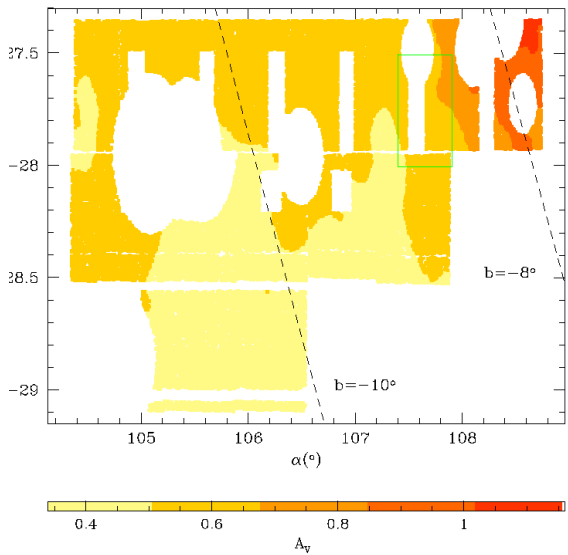}
  \caption{Color-coded map of the extinction $A_V$ in the observed region, from Schlegel et al. (1998). Two lines of constant galactic latitude are shown as reference. Some regions have been eliminated because of very bright stars or bad regions of the CCDs. The rectangle has been observed by Mart\'{\i}nez-Delgado et al. (2004)
.}
  \label{fig:mapaAv}
\end{figure}

Aperture photometry was performed using standard \emph{IRAF} routines, obtaining bright and faint V limiting magnitudes 13.9 and 19.5 respectively. At these low galactic latitudes the extinction is highly variable (Figure \ref{fig:mapaAv}). Magnitudes were corrected using the $E(B-V)$ provided by the Schlegel, Finkbeiner and Davis (1998) dust maps.

\section{RR Lyrae star search}

\begin{figure}[!th]\centering
  \includegraphics[angle=270,width=0.8\columnwidth]{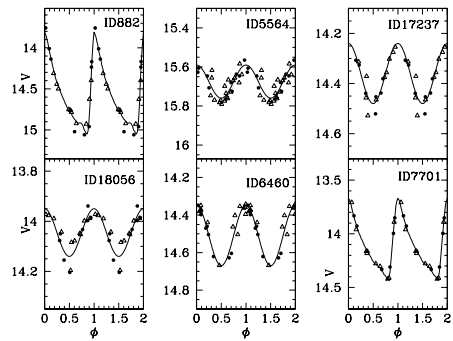}
  \caption{Light curves of confirmed RR Lyrae stars. \\($\bullet$): QUEST data. ($\triangle$): SMARTS data.}
  \label{fig:lcs}
\end{figure}

\begin{figure}[!th]
  \includegraphics[bb=50 400 600 550,clip,width=\columnwidth]{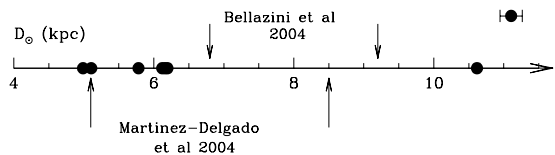}
  \caption{Heliocentric distance distribution of confirmed RR Lyrae stars. Distance estimates obtained by Bellazzini et al (2004) and Mart\'{\i}nez--Delgado et al (2004), are shown between arrows.}
  \label{fig:dist}
\end{figure}

The search for RR Lyrae stars was made among variable stars selected through a $\chi^2$ test (Vivas et al 2004). Then the search was restrained to stars with V magnitudes brighter than 16.2 only, which corresponds to a maximum distance of 13.5kpc. Those are the stars likely to be related to the Canis Major overdensity, according to the distance estimates by Mart\'{\i}nez--Delgado et al (2004) and Bellazzini et al (2004). First, amplitude and color restrictions were applied. Then following the method devised by Layden (1998), \emph{ab} and \emph{c} type light curve templates were fitted to the observational data of the remaining stars, and for each star the four best fitted light curves of each type were visually inspected. Finally, for the best candidates additional observations were obtained with the 1.0 and 1.3m telescopes of the SMARTS consortium at the Cerro Tololo Interamerican Observatory (CTIO). This way 6 RR Lyrae stars were confirmed, having a mean number of $19$ observations each (Figure \ref{fig:lcs}). Of these, 5 lie between 5 and 7 kpc (Figure \ref{fig:dist}), having a mean distance of 5.6kpc and standard deviation 0.5kpc.

\section{Color--Magnitude Diagram}

\begin{figure*}[!t]\centering
  \includegraphics[width=0.85\textwidth]{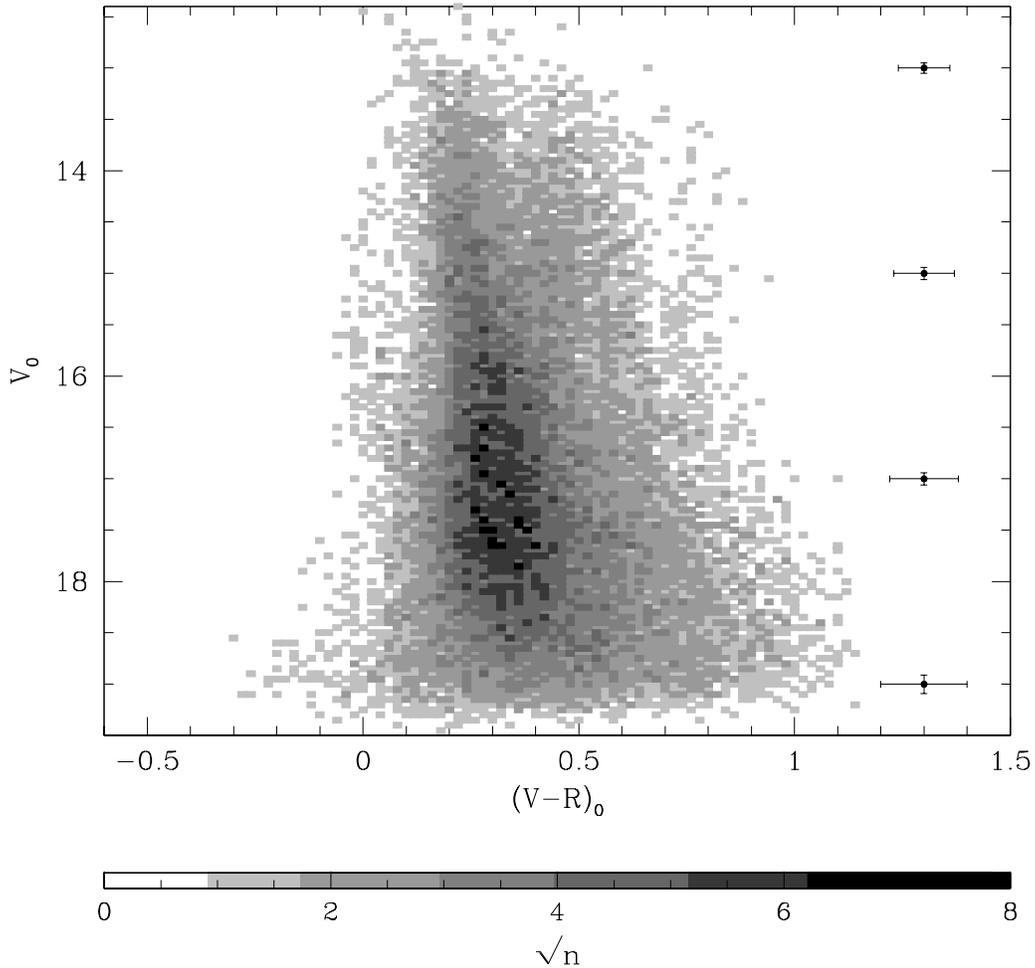}
  \caption{Hess diagram containing about $55.000$ stars. The binning is $0.02$ in $(V-R)_o$, and $0.05$ in $V_o$.}
  \label{fig:Hess}
\end{figure*}

Figure \ref{fig:Hess} shows the Hess diagram of the surveyed region, having $\sim55.000$ stars. Although a weak red giant branch can be identified going from $(V-R)_o\sim0.4,V_o\sim16.1$ to $(V-R)_o\sim0.5,V_o\sim13.6$, we do not detect either a horizontal branch, red clump or main sequence turnoff. According to the distance estimates of Mart\'{\i}nez--Delgado et al (2004), the main sequence turnoff should be close to our limiting V magnitude. The substraction of a control field is intended to be done in the future, in order to be able to perform further analisys of the Hess diagram.

\section{Conclusions} 

The resulting Hess diagram does not show any prominent features that may indicate the clear presence of a stellar population different from the galactic disk. However, at such low galactic latitudes, the contamination from disk stars may hide the presence of any special feature, particularly if its surface density is very low, as is expected in the case that the Canis Major overdensity is indeed a dwarf galaxy in an advanced stage of disruption.

The results yielded by the RR Lyrae search do not show a very high number of such stars. However, between 5 and 7 kpc, 5 confirmed RR Lyrae stars were found with a mean distance of $5.6$kpc. Though not very high, this number of RR Lyrae stars is still higher than expected in that volume of the galactic halo ($\sim1$ RR Lyrae expected). A radial velocity study is being made to confirm whether these stars are physically associated to one another an to the Canis Major overdensity.

\acknowledgements The Llano del Hato Observatory is operated by CIDA for Ministerio de Ciencia y Tecnolog{\'\i}a of Venezuela.

\end{document}